\documentclass[sigchi]{acmart}

\setcopyright{none}
\renewcommand\footnotetextcopyrightpermission[1]{} 

\usepackage{balance} 
\usepackage{multirow}

\begin{document}

%
\title{New advances in the automation of context-aware \\ information selection and guided model assembly}
\renewcommand{\shorttitle}{New advances in the automation of model assembly}

%




\author{Yasmine Ahmed$^1$, Adam A. Butchy$^2$, Khaled Sayed$^1$, Cheryl Telmer$^3$, Natasa Miskov-Zivanov$^1$$^2$$^4$}
\affiliation{%
 \institution{$^1$Department of Electrical and Computer Engineering, $^2$Department of Bioengineering,$^4$Department of Computational and Systems Biology, University of Pittsburgh, $^3$Molecular Biosensor and Imaging Center, Carnegie Mellon University}}
\email{{YAA38,Adam.Butchy,KSS60,NMZivanov}@pitt.edu, Ctelmer@cmu.edu}

\renewcommand{\shortauthors}{Y.Ahmed, et al.}

\maketitle


\section{Introduction}
\begin{figure}[!thb]
\centering
	\includegraphics[width= \linewidth]{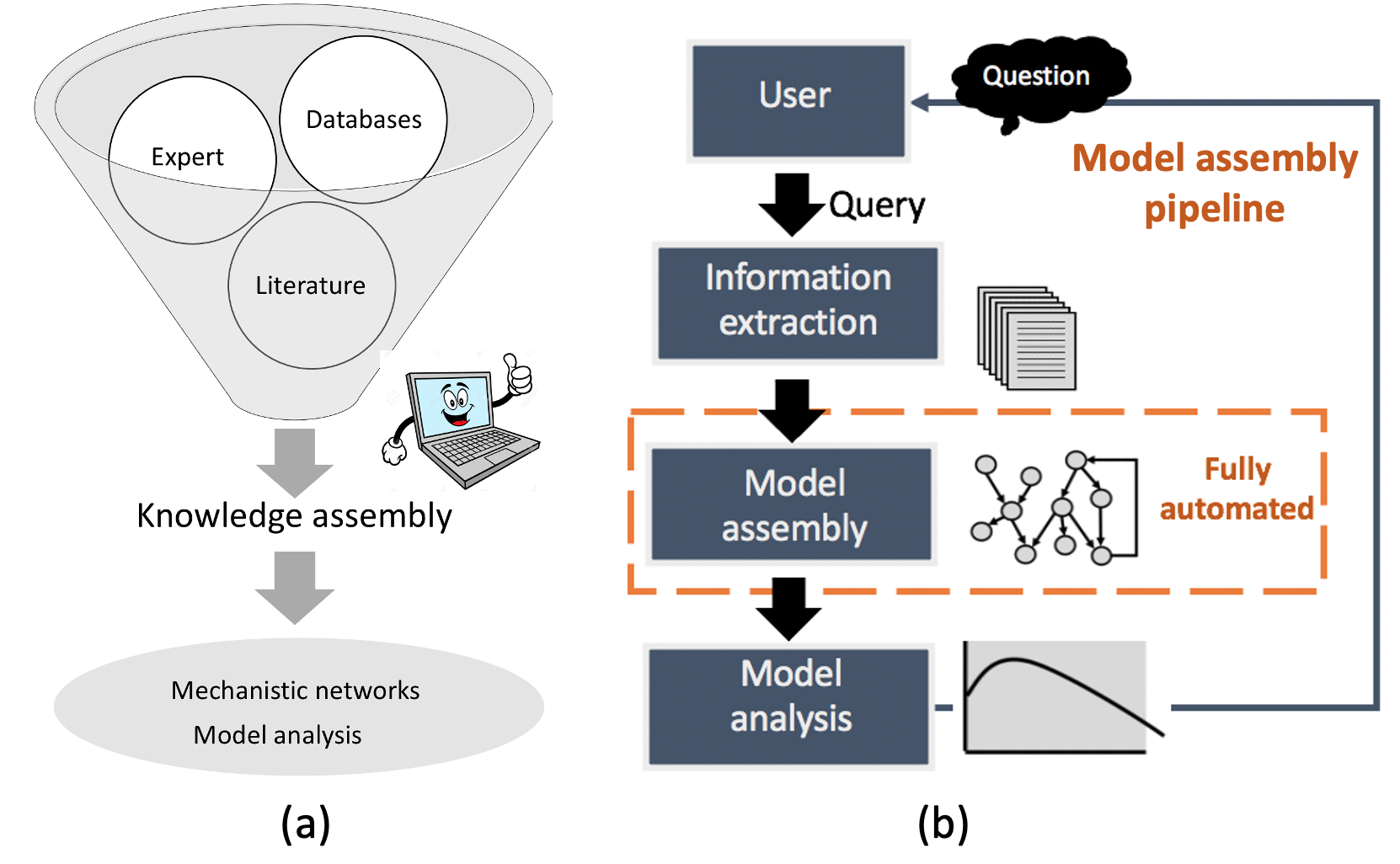}
    \caption{(a) Knowledge assembly, conceptual overview; (b) Automated model assembly pipeline.}
\label{fig:fig1}
\end{figure}

 Modeling complex systems or extending existing models with new information enables a better understanding of these systems ~\cite{Fisher2007}. New information can be extracted from different knowledge sources—such as expert knowledge, published literature and pathway databases—and used to assemble or extend models (Figure ~\ref{fig:fig1} (a)). However, modeling is a time and labor-intensive task, often limited by the knowledge and experience of the modelers. With new research articles published each day, there is a pressing need for an automated method that updates models with new information efficiently and automatically, while preserving the usefulness and accuracy of the original models. 
Recently, there has been a push in the field of synthetic biology to automate the entire pathway of model assembly, starting with collecting biological interactions, assembling a model, and performing simulations ~\cite{Appleton2017}.  A typical model assembly pipeline (Figure ~\ref{fig:fig1} (b)) begins with a question about the system under study. This question is converted into a search engine query to identify and extract the most relevant papers. Biological events are extracted from those papers and used to assemble or extend models. The newly assembled models are then analyzed and evaluated to determine if they satisfy desired system behavior. In this work, we survey the most recent automated model assembly efforts. Specifically, we will review five tools: Layer-based ~\cite{Liang2017}, Genetic Algorithm (GA) based ~\cite{Sayed2018}, ACCORDION ~\cite{ahmed2020accordion}, CLARINET ~\cite{Ahmed2021} and FIDDLE ~\cite{Butchy2021}. We will emphasize the applicability and benefits of each tool using a case study of T cell differentiation model ~\cite{Hawse2015} ~\cite{MiskovZivanov2013} .

\section{Background}

Cellular signaling pathways can be modeled as directed graphs, with nodes representing pathway elements, and edges representing interactions between elements. To study the dynamics of such systems, all the presented tools use executable models, where discrete variables represent states of model elements, and each element can have a state transition function or update function. A baseline model, and an output from a machine reading engine are the inputs to the model assembly pipeline. Each model assembly pipeline generates candidate models of the system under study. Model checking is then used to verify whether each candidate model satisfies a set of properties describing expected behavior of the system. Here, we compare the tools using the same T cell model described in ~\cite{Hawse2015}, and suggested set of interactions from an open-source reading engine, REACH ~\cite{ValenzuelaEscrcega2018}. We expressed both the model and the reading output using an element-based BioRECIPES format ~\cite{Sayed2017} and we used the DiSH simulator ~\cite{8247841} to observe dynamic behavior of the baseline and newly assembled models. We also used statistical model checking ~\cite{Wang2016} ~\cite{Jha2009} ~\cite{Zivanov2013} ~\cite{Zivanov2016} to test all generated models against formally defined properties.
\section{Automated Model Assembly}

\textbf{3.1 Layer-based approach.} In ~\cite{Liang2017}, the authors proposed a method that starts with a baseline model and selects interactions extracted from published literature automatically. The proposed method groups the information extracted from literature into layers, based on their proximity to the important elements in the baseline model. The pieces of information organized in such layers are then added to the baseline model, so that the extended model satisfies predefined system properties. The proposed method helps identify some new interactions without trying the extracted interactions all at once or one interaction at a time. Since the extension method adds new interactions based on their proximity to existing models, this method becomes impractical with large-scale models.\\ \textbf{3.2 GA-based approach.} Another model extension method that uses a Genetic Algorithm ~\cite{Whitley1994} was proposed in ~\cite{Sayed2018}. The authors in ~\cite{Sayed2018} removed a group of elements from an existing model in a random way and they mixed them with randomly created interactions to mimic the output of machine reading engines. Eventually, they applied the genetic algorithm to search for the extensions that optimally reconstructed the model. It has been proved in ~\cite{Sayed2018} that the GA-based approach was able to extract a set of extensions that led to the desired system behavior. The main disadvantages of the GA-based approach include non-determinism, as the solution may vary across multiple algorithm executions on the same inputs, as well as issues with scalability.\\\\ \textbf{3.3 ACCORDION} (\textbf Automated \textbf Clustering \textbf Conditional \textbf On \textbf Relating \textbf Data of \textbf Interactions t\textbf O a \textbf Network).  A tool that automatically and efficiently assembles the information extracted from available literature into models, evaluates the dynamic behavior of newly assembled models, and selects the most suitable model to address user questions as described in ~\cite{ahmed2020accordion}. In contrast to ~\cite{Liang2017} and ~\cite{Sayed2018}, ACCORDION focuses on identifying clusters of strongly connected elements in the newly extracted information that have a measurable impact when added to the model. ACCORDION uses Markov Clustering algorithm (MCL) ~\cite{van2000graph}, an unsupervised graph clustering algorithm, to group interactions obtained from literature by machine reading. Eventually, it finds return paths that start in the baseline model, go through one or more clusters, and end in the baseline model; the baseline model and the clusters on such return path form a candidate model. 
\\\\ \textbf{3.4 CLARINET} (\textbf C\textbf L\textbf A\textbf R\textbf Ifying \textbf N\textbf E\textbf Tworks). Recently, a novel methodology was proposed in ~\cite{Ahmed2021} to expand dynamic network models using the information extracted from published literature by machine reading engines. CLARINET organizes the extracted events as a collaboration graph and uses several novel metrics for evaluating these events individually, in pairs, and in groups. The metrics introduced by CLARINET are based on the frequency of occurrence and co-occurrence of events in literature, and their connectivity to the baseline model. CLARINET is scalable; its average runtime is at the order of seconds when processing several thousand interactions. \\\\\textbf{3.5 FIDDLE} (\textbf Finding \textbf Interactions using \textbf Diagram \textbf Driven mode\textbf L \textbf Extension). A tool described in ~\cite{Butchy2021} that employs two methods based on network search algorithms—Breadth First Addition (BFA) and Depth First Addition (DFA)—to automatically assemble or extend models with the knowledge extracted from published literature. FIDDLE is able to refine models by systematically adding known biological interactions into intermediate models, measuring changes in model performance, and then adding or discarding interactions based on whether they improve the model performance metric. Both BFA and DFA scale linearly with the size of the model they are tasked to extend, and the number of potential interactions with which to extend the model. 

\section{Results and Discussion}

To demonstrate the accuracy, efficiency, and utility of each tool, we have selected a computational model of T cell differentiation ~\cite{MiskovZivanov2013}. Our main goal with this case study is to show that each tool is able to automatically assemble and extend an existing published model into another published and manually built model using new elements and new interactions automatically extracted from published literature. As the final golden model, we used the T cell model published in ~\cite{Hawse2015} and the set of desired system properties discussed in ~\cite{MiskovZivanov2013} and ~\cite{Hawse2015}. The complete list of 27 properties is shown in Table ~\ref{tab:propertyTable}. The golden model and the properties are used to evaluate the automatically assembled model obtained by each tool. Figure ~\ref{fig:fig2} highlights the differences between the results obtained for each tool when tested using statistical model checking. 
The GA-based method features the best performance as scored through statistical model checking. Due to its iterative nature, the time required to perform GA-based extension increases with the number of possible extensions, and can be prohibitively long when applied to large scale models ~\cite{Ahmed2021}. Both ACCORDION and CLARINET balance performance with scalability and can be applied to large scale models, as well as large scale machine reading output, as demonstrated in ~\cite{ahmed2020accordion} and ~\cite{Ahmed2021}. CLARINET scores the newly extracted events based on both the evidence from literature and the connectivity to the baseline model. If the user is interested in collecting new interactions that are strongly connected to each other and strongly connected to the baseline model, then, ACCORDION would be a better choice; since it adds paths of connected interactions, which are at the same time connected to the baseline model. 
The layer-based and BFA methods perform similarly, despite adding different number of extensions to the baseline model. The layer-based method is meant to be applied when the user is interested in collecting new, relevant interactions that are directly connected to the baseline model. The DFA method performed the worst, scoring below the baseline model. This can be attributed to optimizing a scoring metric different than statistical model checking. In fact, both FIDDLE methods attempt to optimize the same metric with the fewest number of extensions to the baseline model. Their poor performance points to their metric being a poor stand in for statistical model checking, and the stipulation to minimize the number of additional extensions as an unnecessary restraint. 
\section{Conclusion and Future work}
Automatically extending models with the information published in literature allows for rapid collection of the existing information in a consistent and comprehensive way. It also facilitates information reuse and data reproducibility. In this review, we described five recent efforts in this direction. We demonstrated the respective benefits and drawbacks of each tool and we tested them on a previously published biological model. These methods and software tools represent a novel effort to replace hundreds or thousands of manual experiments, and have a potential to significantly accelerate the advancement of scientific knowledge. As our future work, we will conduct a more in-depth comparison of the five tools to even more precisely evaluate their advantages and drawbacks. We plan to apply the proposed methods on several other models in different biological domain, and we will work on parallelization of the model checking algorithm to further increase its execution efficiency.

\section{Acknowledgement}
This project was funded by DARPA award W911NF-17-1-0135.

\begin{figure}[!thb]
\centering
	\includegraphics[width= \linewidth]{ure1.png}
    \caption{Comparison of the model checking probability estimates \textit{p} for the baseline model, golden model, and the best model obtained from each of the five tools: Layer-based, GA-based, ACCORDION, CLARINET and FIDDLE (BFA and DFA), when run on a 3.3 GHz Intel Core i5 processor. In the last three rows, we show the number of properties with probability estimates >0.85, the length of time for each method, and the number of extensions added to the baseline model.}
\label{fig:fig2}
\end{figure}

\begin{table}[]
\caption{Set of properties that are observed to be true in T cells ~\cite{Hawse2015} ~\cite{MiskovZivanov2013}.}
\begin{center}
\resizebox{\columnwidth}{!}{%
\begin{tabular}{@{}cl@{}}
\toprule
\multicolumn{1}{l}{Prop\#} & \multicolumn{1}{c}{Description}                                            \\ \midrule
\multicolumn{2}{c}{Scenario 0: No TCR}                                                                  \\ \midrule
1                          & Once deactivated, AKT will remain inactive until end of analyzed period    \\
2                          & Once activated, PTEN will remain active until end of analyzed period       \\
3                          & Once deactivated, FOXP3 will remain inactive until end of analyzed period  \\
4                          & Once deactivated, IL2 will remain inactive until end of analyzed period    \\
5                          & Once deactivated, CD25 will remain inactive until end of analyzed period   \\
6                          & Once deactivated, STAT5 will remain inactive until end of analyzed period  \\
7                          & Once deactivated, mTOR will remain inactive until end of analyzed period   \\
8                          & Once deactivated, mTORC2 will remain inactive until end of analyzed period \\
9                          & Once activated, FOXO1 will remain active until end of analyzed period      \\ \midrule
\multicolumn{2}{c}{Scenario 1: Low TCR}                                                                 \\ \midrule
10                         & Once deactivated, AKT will remain inactive until end of analyzed period    \\
11                         & Once activated, PTEN will remain active until end of analyzed period       \\
12                         & Once activated, FOXP3 will remain active until end of analyzed period      \\
13                         & Once deactivated, IL2 will remain inactive until end of analyzed period    \\
14                         & Once activated, CD25 will remain active until end of analyzed period       \\
15                         & Once activated, STAT5 will remain active until end of analyzed period      \\
16                         & Once activated, mTOR will remain active until end of analyzed period       \\
17                         & Once activated, mTORC2 will remain active until end of analyzed period     \\
18                         & Once activated, FOXO1 will remain active until end of analyzed period      \\ \midrule
\multicolumn{2}{c}{Scenario 2: High TCR}                                                                \\ \midrule
19                         & Once deactivated, AKT will remain inactive until end of analyzed period    \\
20                         & In developing Th, PTEN decreases and remains absent                        \\
21                         & Once deactivated, FOXP3 will remain inactive until end of analyzed period  \\
22                         & Once activated, IL2 will remain active until end of analyzed period        \\
23                         & Once activated, CD25 will remain active until end of analyzed period       \\
24                         & Once activated, STAT5 will remain active until end of analyzed period      \\
25                         & Once deactivated, mTOR will remain inactive until end of analyzed period   \\
26                         & Once activated, mTORC2 will remain active until end of analyzed period     \\
27                         & Once activated, FOXO1 will remain active until end of analyzed period      \\ \bottomrule
\end{tabular}}
\end{center}
\label{tab:propertyTable}
\end{table}

\bibliographystyle{acm}
\bibliography{abstract}

\end{document}